# Deterministic Iteratively Built KD-Tree with KNN Search for Exact Applications (September 2020)

A. Naim *Member, IEEE*, J. Bowkett *Member, IEEE*, S. Karumanchi, P. Tavallali, B. Kennedy

**Abstract**—K-Nearest Neighbors (KNN) search is a fundamental algorithm in artificial intelligence software with applications in robotics, and autonomous vehicles. These wide-ranging applications utilize KNN either directly for simple classification or combine KNN results as input to other algorithms such as Locally Weighted Learning (LWL). Similar to binary trees, kd-trees become unbalanced as new data is added in online applications which can lead to rapid degradation in search performance unless the tree is rebuilt. Although approximate methods are suitable for graphics applications, which prioritize query speed over query accuracy, they are unsuitable for certain applications in autonomous systems, aeronautics, and robotic manipulation where exact solutions are desired. In this paper, we will attempt to assess the performance of non-recursive deterministic kd-tree functions and KNN functions. We will also present a "forest of interval kd-trees" which reduces the number of tree rebuilds, without compromising the exactness of query results.

——————————— ◆ ———————————

## 1 INTRODUCTION

THE interest in KKN search arises from this algorithm's ability to efficiently query multidimensional data and provide such query results as training input to locally weighted learning systems, specifically Locally Weighted Regression (LWR) for function estimation[6]. The kd-tree data structure can improve KNN search from a brute force search, that takes $O(n)$ time complexity to $O(\log n)$, by splitting the search space recursively. The kd-tree data structure is a generalization of binary trees for dimensions $D>2$ and KNN finds $n$ closest points to a query point $X_q$ in multidimensional space [1],[11],[12]. Formally, given $n$ points $P = \{p_1, p_2, p_3, \ldots p_n\}$, $p_i \in \mathbb{R}^D$ and a query point $X_q \in \mathbb{R}^D$, a KNN search finds the k-nearest points of $X_q$ in $P$ [1]. Current open-source software packages do not adhere to quality and safety standards required by aeronautics labs such as JPL (Jet Propulsion Laboratories) or standards set for vehicle software outlined by MISRA (The Motor Industry Software Reliability Association) [7], [18]. For example, these open-source packages over-rely on recursive functions calls, which may cause memory-related "crashes" [7]. Other issues in existing implementations include dynamic memory allocation and deallocation after program initialization, which may cause memory leaks, and ultimately lead to unresponsive systems [7]. Such software "crashes" are catastrophic for mission critical artificial intelligence applications in aerospace, robotics, and autonomous systems. In addition, existing implementations use of probabilistic algorithms, rather than deterministic algorithms, in order to speed up KNN query time and kd-tree build time [1]. These probabilistic algorithms make software verification difficult, which is not desired in applications with stringent safety requirements. Many Approximate Nearest Neighbor (ANN) implementations use an approximation error tolerance $\varepsilon$ [12]. Therefore, ANN's technique is to calculate the distance between the query point $X_q$ and nearest neighbor $X_{approx}$ within a factor of $1 + \varepsilon$ of the distance to the exact nearest neighbor $X_{exact}$, which means $d(X_q, X_{approx}) <= (1 + \varepsilon) * d(X_q, X_{exact})$ [12]. One can notice some of $X_q$'s exact neighbors that are within a factor of $1+\varepsilon$ of the distance $D_k$ of $X_q$'s approximate kth-nearest neighbor could be missed [12]. Therefore, our challenge is to implement a minimum viable kd-tree and exact KNN library, that adheres to the JPL C coding standard [7], and to explore novel ways of improving kd-tree and KNN run times without compromising exactness or determinism.

## 2 KD-TREE AND KNN

Typical kd-tree algorithms build kd-trees by partitioning input points [1],[11],[12]. Kd-tree add, search, delete operations are all extensions of kd-tree traversal and not much different than their binary tree relatives, refer to Algorithm 1. The guarantee of near $O(\log n)$ traversal is based on some "splitting" criteria, which starts comparison between the root of the tree vs "splitting" criteria, and traverses based on that criteria either left, right, the tree until some node is found or not. Usually splits are based on a descriptive statistic such as median or variance of the node values in the tree, with the ultimate goal of balancing the tree for efficient add, delete, search operations. We recorded ten iterations of KNN search on a multicore Intel Core x86 architecture laptop, to retrieve 30 neighbors of a query point in a kd-tree of 60,000 points. The

————————————————

*The authors are with the Jet Propulsion Laboratory, California Institute of Technology, 4800 Oak Grove Drive La Cañada Flintridge, CA 91011. E-mail: aryan.e.naim@jpl.nasa.gov.*

22

results suggested that on average unbalanced trees of that size were 23% slower than balanced trees. There are many different implementations of kd-trees, we will explore one of the most prevalent designs [11], [12]:

I. Cycling through *D* dimensions – partition based on a median of dimension d1, then median of dimension $d_2,…,d_k$ before cycling back to dimension d1[11]. Repeat until the tree is built or perform other traversal operations.

There are also other variations such as Quadtrees, Octtrees, BSP-trees, and R-Trees to say the least, for more detailed information on these other kd-tree variations refer to (Samet, 2006) [12].

## 3 DATA-STRUCTURE AND HEAP MEMORY MANAGEMENT

The main data-structure is represented in *Figure. 1 . UML diagram of kd_tree's and kd_tree_node.*, which is titled "kd-tree". This data-structure has zero to many nodes of type "kd_tree_node". The kd_tree encapsulates the root node of the tree. The kd_tree_node struct has a left pointer to a child leaf, a right pointer to child leaf, a parent pointer to a parent of the same type, and pointer to the actual multi-dimensional data of length *D*, and finally, the attribute titled "distance_to_neighbor" that holds the distance to nearest when a collection of called "kd_tree_nodes" is returned as result to a KNN search.

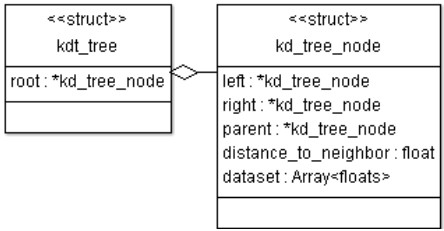

Figure. 1 . UML diagram of kd_tree's and kd_tree_node.

Now that we have presented the node and tree data structures, let's turn our attention to memory management. In order, to minimize the risk of "memory leakage" that is when dynamically allocated memory, is not deallocated or "freed", there are recommended software design patterns. These design patterns prohibit memory allocation and deallocation after program initialization, as per JPL Institutional Coding Standard Rule 5[7]. These guidelines significantly affected our software implementation and limited our algorithmic options. The design pattern for such stringent memory management requires a single memory allocation function, in our case called "kdtree_alloc" that dynamically allocates all memory required to build the kd_tree at program start and only called once during the entire duration of the program. Furthermore, an initialization function is called "kdtree_init" that sets initial values and is used to recycle allocated memory in case the program needs to run many cycles. Finally, "kdtree_free" deallocates all kd_tree memory that was allocated and again only called once at the end of the program. This design pattern reduces the chance of "memory leakage", because memory allocation and deallocation are restricted to the start and end of the program respectively. The memory management technique above is widely used in embedded software environments. Furthermore, static code analysis tools and memory profiling tools such as Valgrind C [13] should be used to eliminate memory-related bugs.

## 4 TRAVERSALS AND KNN

At the heart of all the iterative kd-tree algorithms one will see a common traversal logic, refer to code snippet Algorithm 1. Algorithm 1 is not a full procedure; however, building block which may be used to add, point search, deletion, or traverse trees. If the root is not empty then the procedure traverses the tree hierarchy, by either going left or right based on a comparison criterion, in our case the median of a particular dimension compared to input data, lines 5-11. Finally, once the right node is encountered the data is either added, returned, or deleted depending on particular operation.

**Algorithm 1.** *An O (log n) iterative algorithm for traversing a kd-tree based on input data*
**Procedure** kd_tree_add_record(root, key [], k_dimensions)
**Input:**
root is pointer to root node,
key [] is array of data to be inserted,
k_dimensions - dimensions of the data
**Output:** void
   1   median = 0, cd = 0
   2   current → root
   3    **while** current is NOT empty
   4      parent = current
   5      cd = depth % k_dimensions
        /*median- splitting criteria*/
   6      median=**procedure** kd_tree_get_column_median(cd)
   7        **if** (key[cd] < median) **then**
   8          current = current.left /*go left*/
   9        **else**
 10          current = current.right /*go right*/
 11      depth++
 12   **end while**
 13   **end procedure**

Below is the crucial rebuild algorithm that ensures on average *O (log n)* traversals, however the rebuild procedure itself has *O (n)* complexity.

**Algorithm 2.** *O(n) complexity kd-tree rebuild routine which is used to create a balanced tree based on new median*

*calculations.*
**Procedure** kd_tree_rebuild (root, k_dimensions)
**Input:**
root – root node of kd-tree
k_dimensions - dimensions of the data
**Output**: Boolean 0 or 1
1. allow only read operations for entire library
2. in_order_array ← **procedure** traverse (root)
3. result_size = size of in_order_array
4. extract column wise values of in_order_array into columns_median_processing_space heap
5. **if** median algorithm is textbook median **then**
6.    **procedure** insertion_sort(columns_median_processing_space, result_size)
7. **end**
8. **else** /*do nothing*/
9. **end**
10. **procedure** median(columns_median_processing_space,result_size)
11. calculated median values ← columns_median_processing_space
12. batch delete kd-tree node heap
/*insert previous nodes 1 by 1 to rebuild*/
**for** i=0 **to** i< result_size
   **procedure** kd_tree_new_node (result_size[i], k_dimensions)
13. Unlock read only operations for entire library.
/*if any errors occur along the way, set result_size=-1*/
14. **return** result_size
15. **end procedure**

There are two popular versions of the KNN search algorithm, one version attempts to return nearest neighbors based on a desired number of neighbors from a query point $X_q$ which is represented by the pseudocode in Algorithm 3. Another version of KNN attempts to return nearest neighbors within the desired radius; were $X_q$ is the center of that radius. The pseudocode in Algorithm 3 is inspired by (Samet, 2006) [12].

**Algorithm 3.** *Attempts to return nearest neighbors based on a desired number of neighbors from a query point $X_q$.*
**Procedure** kd_tree_knn (root, data[],k_dimensions)
**Input:** root is pointer to root node
data_point [] – is the query point
k_dimensions- dimensions of the data
number_of_nearest_neighbors – requested number of neighbors
**Output**: number of nearest neighbors found
kd_tree_knn(root, const float data_point[],
const int k_dimensions,
int number_of_nearest_neighbors)

1. nearest_counter = 0 , heap_index = 0
2. current → NIL
3. cd = 0, depth = 0
4. median = 0.0, distance = 0.0
5. **if** root != NIL **then**
   Initialize node_knn_result_space heap
6.    current → root
/* start traverse*/
7.    **while** current is NOT empty
8.      distance= **procedure** kd_tree_n_dimensional_euclidean(
9.      data_point,
10.      current.dataset,
11.      k_dimensions)

/* simulate priority queue via array*/
/*if array is NOT full*/
12.     **if** (heap_index < procedure kd_tree_get_rows_size()) **then**
/*insert in array*/
13.  node_knn_result_space[heap_index].dataset = current.dataset
14. node_knn_result_space[heap_index].distance_to_neighbor = distance
15.    heap_index++
16.   **end**
17.   **else**
18.    **procedure** insertion_sort_based_on_distance(node_knn_result_space,
19.    nearest_counter)

/*is the smallest distance greater than current*/
20.    **if** (node_knn_result_space[0].distance_to_neighbor >distance) **then**
/*insert in array at the top, therefore index=0*/
21.     node_knn_result_space[0].dataset = current.dataset
22.     node_knn_result_space[0].distance_to_neighbor = distance
23.    **end**
24.   **end**

/* Calculate dimension of comparison */
25.    cd = depth % k_dimensions

```
26      median = procedure
        kd_tree_get_column_median(cd)
27      if (data_point[cd] < median) then
28          current → current.left
29      end
30      else
31          Current → current.right
32      end
33      depth++
34   end while
     /*end travrse */
35   nearest_counter = heap_index
36   if (nearest_counter > 0) then
37       procedure sort (node_knn_result_space,
38       nearest_counter)
39   end
40   end
41   return number_of_nearest_neighbors
42 end procedure
```

## 5 EXPERIMENTAL RESULTS

The primary drawback of KNN search, is speed when compared to ANN. Kd-trees are used to efficiently traverse multi-spatial data in $O(\log n)$ time, this is only guaranteed when the kd-tree is balanced. Kd-trees organize data with relatively high dimensions (D>20), with significant rebalancing or rebuild costs due to the "curse of dimensionality". This performance does not meet any latency requirements for exact application such real-time image processing for aeronautics, robotics, autonomous systems that require low latency and sub 1 second response times. On an embedded system with less processing power, the performance will further degrade. Over the years there has been fair amount of research to speed up KNN search algorithm directly by relaxing criteria of exact neighbors returned with approximate neighbors which is acceptable for many applications but not all applications [1], [12], [14], [15]. Of the two problems KNN search speed and tree build speed, we will attempt tackle the later problem with goal of minimizing rebuilds. We present a series of benchmarks as represented in Fig. 2 to Fig. 5 on virtualized Linux OS on top of a modern multicore Intel x86 core architecture laptop. Comparing total kd-tree build times of trees with different sizes represents a geometric rise in tree build time, despite the only linear growth in number of nodes from 1,000 to 12,000 nodes. Fig 2. shows that a simple polynomial regression fitted to tree build times looks quadratic. Trees become unbalanced approximately when the number of nodes doubles [1]. The tree rebuild procedure seems to be the main contributor to total tree build time. Fig. 3 and Fig. 4 will compare tree rebuild times for every time the tree rebuilds versus the current number of nodes in the tree at that time. We recorded average rebuild times over 10 iterations for thresholds 2 and 3. Threshold's 2 or 3 refer to when the software decides to trigger a rebuild when the tree doubles or triples in size. In Fig. 3 and Fig. 4 we can see that initial rebuilds have negligible cost, however suddenly the cost of each rebuild grows. We will now compare the cost of KNN search of self-balancing kd-tree Fig. 5, which re-balances or re-builds every time the tree doubles in size, hence a rebuild threshold of 2 versus KNN search in a unbalanced tree. Fig. 5, presents average execution times for KNN searches were 30 nearest neighbors were performed for an arbitrary query point $X_q$. Tests were run for kd-trees of size 125 to 64,000 nodes. A kd-tree of size 125 nodes found 30 nearest neighbors of $X_q$ returned results within milliseconds, versus a kd-tree of size 64,000 nodes for the same query returned results within 5.46 secs. We performed the same for KNN searches on an unbalanced tree and plotted the results in Figure 6, although graphs look nearly identical, KNN searches starting at kd-tree size 8,000 to 64,000 were from 3% to about 18% slower versus a balanced tree of the same size as the number of nodes grew. We were expecting much slower performance for an unbalanced tree, based on the theoretical time complexity of traversing unbalanced tree with $O(n)$ vs $O(\log n)$ time complexity for a balanced tree.

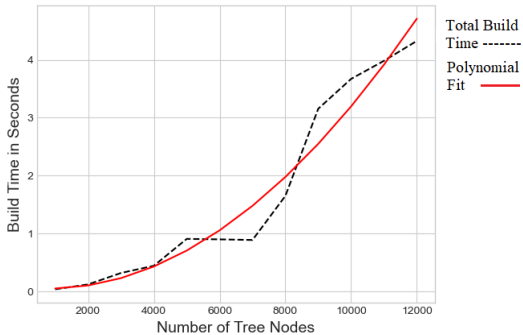

Figure. 2 Total Kd-tree Build Time Per Number of Tree Nodes Threshold=3

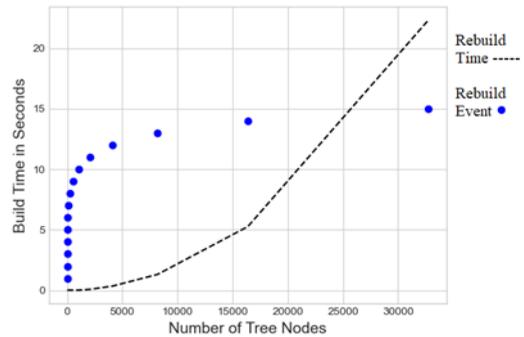

Figure. 3 Total kd-tree Build Time Per Number of Tree Nodes. Build Threshold=2

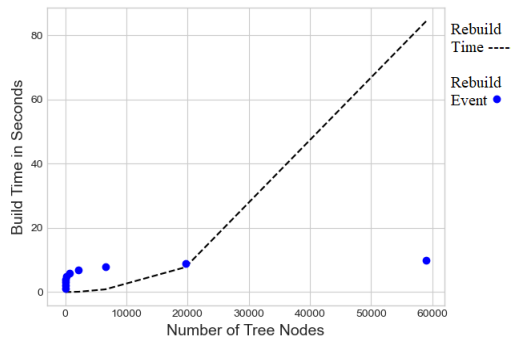

Figure. 4 Total kd-tree Build Time Per Number of Tree Nodes. Build Threshold=3

## 6 FOREST OF INTERVAL KD-TREES

The dilemma can be summarized as a compromise between faster tree builds versus faster KNN searches, and the data suggests that kd-tree rebuilds are a major bottleneck. There seems to be an augmented solution that contains a combination of kd-trees and interval trees [16], hence "forest of interval kd-trees". Therefore, the proposal is to minimize tree rebuilds by splitting a single large tree with $n$ nodes to many smaller trees, using a "divide & conquer" into a forest, and keep track of the range of values within each kd-tree, similar to interval trees. Given a kd-tree with rebuild threshold of $b$, the tree will be rebuilt when (1) is true.

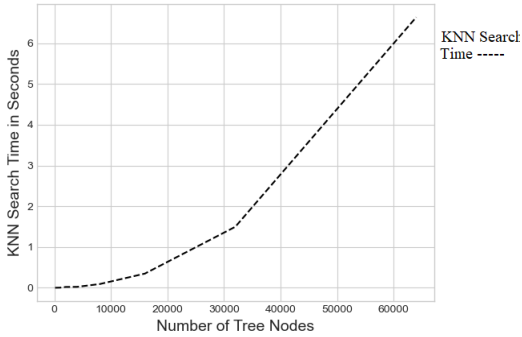

Figure. 5 Total KNN Search Time Per Number of Nodes. Rebuild Disabled

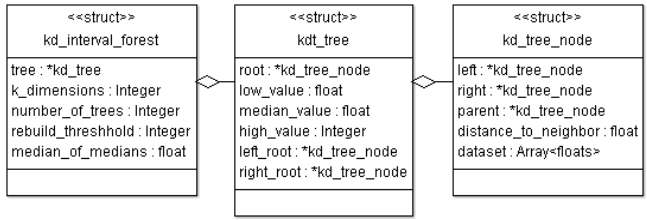

Figure. 6 Architecture #1 UML Diagram.

$$b \geq \frac{number\_of\_nodes_{current}}{number\_of\_nodes_{previous}}$$

Equation 1. Rebuild Threshold

With a default rebuild threshold of $b$=2 and $n$ number of nodes, then the number rebuilds in a single large kd-tree would be approximately (2).

$$r \approx \ln(n)$$

Equation 2.

Therefore, to minimize tree rebuilds via the method of splitting trees, the kd-forest with kd-trees having a $b$=2 rebuild threshold, will have a minimum (3).

$$number\_of\_trees = \lceil \ln(n) \rceil + 1$$

Equation 3.

For example, a single kd-tree with $n$=64,000 nodes would have had ln(64,000) 12 rebuilds, where each subsequent rebuild would take longer to execute than the previous rebuild, refer to Fig. 3 and Fig. 4. However, using "forest of interval kd-trees", the size of each kd-tree can be reduced (4).

$$t_s = n/(number\_of\_trees)$$

Equation 4. Reducing the number of rebuilds by diving a single kd-tree of n nodes to many kd-trees each size $t_s$.

Therefore, reducing the total number of rebuilds for each kd-tree, which in turn will reduce KNN search time due to smaller tree sizes. Finding candidate tree to start the KNN search will be handled by techniques used in interval trees, refer to Algorithm 4 [16]. However, even this reduction in tree size (4) is not enough to meet typical low-latency scenarios for computer vision based artificial intelligence algorithms known as V-SLAM (vision based simultaneous localization and mapping) in autonomous systems. For example, in the case of autonomous vehicles, the average truck speed [19] on American roads is a record as 46 mph or about 74 kmph, in terms of seconds we are looking at 67.5 fps (feet per second) and or 20.6 mps (meters per second). If the autonomous vehicle was using a kd-tree with input from a low-end 640x480 resolution image sensor then tree build time would be somewhere between 63-70 seconds assuming the best-case scenario of an in-vehicle computer having a multicore Intel X86 processor, and that performance is not acceptable given typical vehicle speeds. Furthermore, today's vehicles have cameras with resolutions of near 3849x1929, hence a 7.42-megapixel camera [20]. Therefore, to overcome such shortcomings, we must augment our "forest of interval kd-trees" with parallel processing techniques. Each kd-tree will have to be built and balanced using multiple threads in parallel. We will present 2 architectures for the "forest of interval trees", referred to as Architecture #1 represented by the UML diagram in Fig.6. Architecture #1 Fig. 6 looks intuitive, with similar traversal patterns that were used in the the kd_tree to traverse nodes with a single kd_tree. Architecture #1 is inspired by Binary Search Trees [16]. Kd_interval_forest has a reference to a single kd_tree as the root tree of the forest. The kd_trees can be traversed and updated relative to each other in a forest based on the individual kd_trees mid_value, or median value of its current nodes, refer to Architecture #1 Fig. 6. However, Architecture #1 Fig. 6 has introduced a complexity, due to coupling of each kd_tree with other kd_trees within the forest via left, right pointers. This coupling will need to be managed when each kd_tree is rebalanced in terms of its internal nodes, which has ripple effect in the forest hierarchy. Worst case an entire forest maybe needs to rebalanced. Coupling makes distributed or parallel computing difficult. In order to remove this coupling, we introduce Fig. 7 Architecture #2, and associated Algorithms 4, and 5.

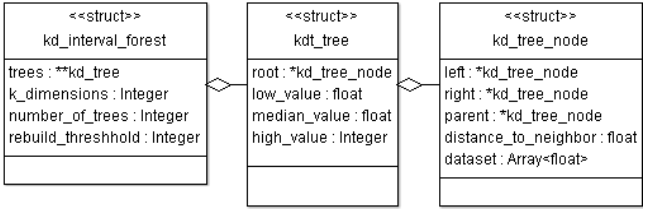

Figure. 7 Architecture #2 UML Diagram.

We introduce Fig.7 Architecture # 2, the kd_interval_forest struct has an array of kd_tree's, and the kd_tree struct has no left_root or right_root pointers, instead, the kd_trees are sorted based on some center value such as the median value. Fig.7 Architecture # 2, no longer requires pointers between trees and a rebalance operation in an individual tree will not require rebalance at kd_tree forest level. After each rebalance operation, the array of kd_tree root pointers that are maintained by kd_interval_forest are sorted based on the median values of each tree. Instead of allocating heap memory for a single tree root as a reference to the rest of the roots in the forest, we allocated memory for the array of tree roots. For efficient forest traversal we must sort the "trees" array every time any new tree is built or when any tree is rebalanced. Architecture #2 frees us from dependency between trees and the potential burden having to swap pointers between kd-trees after each kd-tree is rebalanced. The most important algorithm in a "forest of interval kd-trees", is the algorithm that finds the correct candidate kd-tree within the forest which will then be passed that tree by reference to regular add, delete, point search, or KNN search operation. Refer to "kd_forest_candidate_tree_arch2" Algorithm 4.

**Algorithm 4** — *Find the right tree in the forest time complexity of O (log n).*
**Procedure** kd_forest_candidate_tree_arch2(kd_interval_forest,query_point)
**Input:**
kd_interval_forest - pointer to kd_interval_forest
query_point – point to searched for
**Output:** kd_tree
1. kd_tree_node→NIL
2. current_tree→NIL
3. trees sort kd_interval_forest.trees based on median value
4. /*iterative binary search based on tree median*/
5. start_index = 0
6. end_index = size of kd_interval_forest .trees -1
7. **while** start_index <= end_index
8.    middle = start_index - end_index
    current_tree←kd_interval_forest .trees[middle]

1.   **if procedure** kd_interval_forest_overlap(query_point, current_tree) **then**
2.     /*we found tree that overlaps our query point*/
3.     **break**
    /*for all dimensions*/
    go_left =0
    **for** i=0 **to** i= kd_interval_forest.k_dimensions
4.       **if** kd_interval_forest .trees[middle].median_val < query_point **then**
5.         go_left = 1
6.       **end**
7.     **end for**
8.     **if** go_left **then**
9.       start_index = middle + 1
10.     **end**
11.     **else**
12.       start_index = middle - 1
13.     **end**
14. **end while**
15. **return** current_tree
16. **end procedure**

The KNN search within a forest is not much different than KNN search for single tree. Once the right candidate tree is found, one would pass kd-tree by reference to the KNN search procedure, refer to Algorithm 5.

**Algorithm 5** — *K Nearest Neighbor (KNN) Search within a forest and within the correct tree.*
**procedure** kd_forest_knn (kd_interval_forest,query_point,n_nearest)
**input:**
kd_interval_forest - pointer to kd_interval_forest
data– data to be inserted
**Output**: number of nearest neighbors found
1. n_nearest_found = 0
   /*Refer to Algorithm 4*/
2. kd_tree<- **procedure** kd_forest_candidate_tree_arch2 (kd_interval_forest, query_point)
3. **if** tree !=NIL **then**
/* kd_tree_knn – refer to Algorithm 3*/
   n_nearest_found <- **procedure** kd_tree_knn (tree, point, kd_interval_forest.k_dimensions)
4. **end**
5. **return** n_nearest_found

We have presented the essence of a forest kd_trees that can now traversed in *O (log n)* given that trees are sorted within the heap which takes *O (n log n)* time itself.

## 8 CONCLUSION AND FURTHER WORK

Based on our various tests with varying kd-tree sizes, we assess that naïve deterministic kd-tree implementations may be suitable for robotics applications that have less than 5,000 data points at a time in a single kd-tree, however, naïve deterministic kd-tree is not suitable for near-real-time tasks with larger inputs. The need for accuracy, larger data sets, and less latency; typical in image processing tasks or autonomous systems, will require a "forest of interval kd-trees" approach that will need to be augmented with parallel processing techniques. Future work should implement "forest of interval kd-trees" and associated Algorithms 4 and 5. "Forest of interval kd-trees" should be deployed on embedded boards and benchmarked against existing libraries such as Nanoflann or FLANN.

## ACKNOWLEDGMENT

The research described in this paper was carried out at the Jet Propulsion Laboratory, California Institute of Technology, under a contract with the National Aeronautics and Space Administration. © 2020 California Institute of Technology. Government sponsorship acknowledged.

**Aryan Naim** is a Senior Software Engineer in Mission Control Systems Deep Learning Group at NASA's Jet Propulsion Laboratory (JPL). He recieved Bachelors in Information Technology from The American University in Dubai. Post Graduate studies in Embedded Systems Engineering at University of California Irvine. He has over 12 years of software development experience focusing on mobile, embedded, resource constraint devices.

**Joseph Bowkett** is a Robotics Technologist in the Manipulation and Sampling group of the Mobility and Robotic systems section at NASA's Jet Propulsion Laboratory (JPL). He attained a Bachelor of Engineering (hons) from the University of Auckland, followed by a MS & PhD at the California Institute of Technology having conducted research primarily at JPL. His work focuses on applying functional autonomy techniques to the problem of robotic manipulation and sampling in poorly characterized environments.

**Sisir Karumanchi** joined JPL in 2014 as a Robotics Technologist and is a member of the Mobility and Robotic Systems Section at JPL. Sisir received a Bachelor's degree in Mechatronic Engineering from the University of Sydney in 2005, and completed his Ph.D. in Field Robotics at the University of Sydney in 2010. During 2011-2014, he was a postdoctoral associate in the Robotic Mobility Group at the Massachusetts Institute of Technology. He was the software lead for the JPL entry to the DARPA Robotics Challenge finals.

**Peyman Tavallali** is a Science Data Machine Learning Researcher at NASA's Jet Propulsion Laboratory (JPL) Machine Learning and Instrument Autonomy (MLIA) group. In 2014, he got his PhD, in Applied and Computational Mathematics, from California Institute of Technology. Dr. Tavallali conducts research in theoretical and applied machine learning towards design of automated and adaptive systems.

**Brett Kennedy** is the Group Supervisor of the Robotic Vehicles and Manipulators Group at JPL. Brett graduated from University of California, Berkeley in 1996 with his B.S. in mechanical engineering. In 1997, he received his M.S. in mechanical engineering from Stanford University concentrating on mechatronics and robotics. He was also the team lead for the JPL entry to the DARPA Robotics Challenge.